\documentclass[aps,prl,final,twocolumn,showpacs]{revtex4}

\usepackage{amsmath,amssymb}
\usepackage{graphicx}

\newtheorem{lemma}{Lemma}

\newcommand{\dd}{\,\mathrm{d}}
\newcommand{\ts}{\hspace{0.5pt}}
\newcommand{\nts}{\hspace{-0.5pt}}

\newcommand{\ZZ}{\mathbb{Z}}
\newcommand{\RR}{\mathbb{R}\ts}
\newcommand{\NN}{\mathbb{N}}
  
\begin{document}

\title{Scaling of the Thue--Morse diffraction measure}

\author{Michael Baake$^{a}$, Uwe Grimm$^{b}$ and Johan Nilsson$^{a}$}
\affiliation{{}$^{a}\!$Fakult\"{a}t f\"{u}r Mathematik, 
Universit\"{a}t Bielefeld,
Postfach 100131, 33501 Bielefeld, Germany\\
{}$^{b}\!$Department of Mathematics and Statistics,
The Open University,
Walton Hall, Milton Keynes MK7 6AA, UK}

\begin{abstract}
  We revisit the well-known and much studied Riesz product
  representation of the Thue--Morse diffraction measure, which is also
  the maximal spectral measure for the corresponding dynamical
  spectrum in the complement of the pure point part.  The known
  scaling relations are summarised, and some new findings are
  explained.
\end{abstract}

\pacs{61.05.cc,  
      61.43.-j,  
      61.44.Br  
     }

\maketitle

\section{Introduction}

The Thue--Morse (TM) sequence is defined via the binary substitution
$1\mapsto 1\bar{1}$, $\bar{1}\mapsto \bar{1}1$; see \cite{AS,tao} and
references therein for general background. The corresponding dynamical
system is known to have mixed (pure point and singular continuous)
spectrum \cite{Q,ME,Kea}, with a pure point part on the dyadic points
and a singular continuous spectral measure in the form of a Riesz
product. The latter coincides with the diffraction measure
$\widehat{\gamma}$ of the TM Dirac comb with weights $\pm 1$; compare
\cite{BG08} for details.

The Riesz product representation of the TM diffraction measure reads 
\begin{equation}\label{eq:TM-Riesz}
  \widehat{\gamma}\, = 
  \prod_{n\ge 0} \bigl( 1 - \cos(2^{n+1}\pi k)\bigr),
\end{equation} 
with convergence (as a measure, not as a function) in the vague
topology; see \cite{Z} for general background. The singular continuous
nature of $\widehat{\gamma}$ is traditionally proved \cite{Q,Nat} via
excluding pure points by Wiener's criterion \cite{Wie,Mah} and
absolutely continuous parts by the Riemann--Lebesgue lemma
\cite{Kaku}; compare \cite{BG08,tao} and references therein for
further material. 

Since diffraction measures with singular continuous components do
occur in practice \cite{Withers}, it is of interest to study such
measures in more detail. Below, we use the TM paradigm to rigorously
explore the scaling properties of `singular peaks' in a diffraction
measure, combining methods from harmonic analysis and number theory;
for further results of a similar type, we refer to
\cite{CSM,GL,Zaks,ZPK1,ZPK2} and references therein.

\section{Ergodicity properties}

In what follows, we employ various Birkhoff sums, and implicitly
explore a range of uniform distribution properties, either on the
unit interval $[0,1]$ or on various finite subsets of it; we refer
to \cite{KN} for background.

Let us first observe that the mapping $T$ defined by $x \mapsto 2 \ts
x \bmod 1$ maps $[0,1]$ into itself and leaves Lebesgue measure
invariant. Moreover, $T$ is \emph{ergodic} relative to Lebesgue
measure \cite{HL,Kac}, wherefore Birkhoff's ergodic theorem gives us
the following result.

\begin{lemma}\label{lem:ergodic}
  Let\/ $g \in L^{1} \bigl( [0,1] \bigr)$ and consider the mapping\/
  $T$ defined by\/ $x \mapsto 2 \ts x \bmod 1$. Then, for Lebesgue
  almost every point\/ $k\in [0,1]$, one has
\[
    \lim_{N\to\infty} \frac{1}{N}
    \sum_{n=0}^{N-1} g ( T^{n} (k)) \, = 
    \int_{0}^{1} g(x) \dd x \ts .
\]
\end{lemma}

Consider the function $f$ defined by 
\[ 
   f(x) \, = \, \log\bigl(1-\cos(2\pi x)\bigr)
\]
on $[0,1]$.  It has singularities at $x=0$ and $x=1$, which are both
integrable (via standard arguments), so $f\in L^{1} \bigl([0,1]
\bigr)$. Note that $f$ is not Riemann integrable in the proper sense,
though it is in the generalised sense of an improper integral. Still,
this means that we cannot directly apply uniform distribution without
some additional argument, which is ultimately equivalent to the
approach via Birkhoff sums. In fact, since $f$ has an obvious
extension to a $1$-periodic function on $\RR$, sums of the form
$\frac{1}{N} \sum_{n=0}^{N-1} f(2^{n+1} \pi x)$ can be analysed, for
almost all $x\in\RR$, via Lemma~\ref{lem:ergodic}.

Here, via an explicit calculation, one gets
\begin{equation}\label{eq:int}
   \int_{0}^{1} \log\bigl(1-\cos(2\pi x)\bigr)\dd x
   \, = \, -\log(2)\, .
\end{equation}
We shall also need a discrete analogue of this formula.  Via
$1-\cos(2\vartheta)=2\bigl(\sin(\vartheta)\bigr)^{2}$ together with
the well-known identity $\prod_{m=1}^{n-1} \sin(\pi\frac{m}{n}) =
n/2^{n-1}$, one can derive that
\begin{equation}\label{eq:qsum}
    \sum_{m=1}^{n-1} \log \bigl(1-\cos(2\pi \tfrac{m}{n})\bigr)\, = \,
    \log\biggl(\frac{n^{2}}{2_{\phantom{T}}^{n-1}}\biggr)
\end{equation}
holds for all $n\ge 1$, with obvious meaning for $n=1$.
\smallskip

\section{Riesz product}

A direct path to the Riesz product of the TM diffraction measure can
be obtained as follows. Consider the recursion $v^{(n+1)}= v^{(n)}
\bar{v}^{(n)}$ with initial condition $v^{(0)}=1$, which gives an
iteration towards the one-sided fixed point $v$ of the TM substitution
on the alphabet $\{1,\bar{1}\}$. If we define the exponential sum
\[
    g_{n}(k) \, = \, \sum_{\ell=0}^{2^{n}-1} v^{}_{\ell}\, e^{-2\pi ik \ell},
\]
where $v = v^{}_{0} v^{}_{1} v^{}_{2} \cdots$, the function $g_{n}$ is
then the Fourier transform of the weighted Dirac comb for $v_{n}$,
when it is realised with Dirac measures (of weight $\pm 1$) on the
left endpoints of the unit intervals that represent the symbolic
sequence of $v_{n}$. In particular, one has $g^{}_{0}(k)=1$ and
\[
   g_{n+1}(k) \, = \, \bigl(1- e^{-2\pi ik 2^{n}}\bigr)\, g_{n}(k)
\]
for $n\ge 0$, so that
\[
   \bigl| g_{n+1}(k)\bigr|^{2} \, = \, 2\ts \bigl| g_{n}(k)\bigr|^{2}
   \bigl(1 - \cos(2^{n+1}\pi k)\bigr).
\]
One can then explicitly check that 
\[
   f_{n}(k) \, := \, \frac{1}{2^{n}}\bigl|
    g_{n}(k)\bigr|^{2} = \prod_{\ell=0}^{n-1}
   \bigl( 1 - \cos(2^{\ell+1} \pi k)\bigr) \ts ,
\]
which reproduces the Riesz product of Eq.~\eqref{eq:TM-Riesz} in the
sense that $\lim_{n\to\infty} f_{n} = \widehat{\gamma}$ as measures in
the vague topology.

As $g_{n}$ corresponds to a chain of length $2^{n}$, the growth rate
$\beta (k)$ of the intensity at $k$ (when this rate is well-defined) is
obtained as
\[
    \beta (k) \, = \, \lim_{n\to\infty}
    \frac{\log \bigl(f^{}_{n} (k)\bigr)}{n \ts \log (2)} \ts .
\]
Let us now consider the growth rate for various cases of the wave
number $k$.

\emph{Case A.} When $k=\frac{m}{2^{r}}$ with $r \ge 0$ and $m\in\ZZ$,
all but finitely many factors of the Riesz product \eqref{eq:TM-Riesz}
vanish, so that no contribution can emerge from such values of $k$. In
fact, these are the dyadic points, which support the pure point part
of the dynamical spectrum. They are extinction points for the
diffraction measure of the balanced weight case considered here; 
compare \cite{ME} for a discussion of this connection.

\emph{Case B.}  Since $f$ is a $1$-periodic function, we clearly have
$\beta (k) = \beta( k \bmod 1)$, so we can use Birkhoff sums with
arguments reduced to the unit interval in conjunction with
Lemma~\ref{lem:ergodic}.  Then, for Lebesgue almost all $k\in\RR$, one
obtains the growth rate
\[
\begin{aligned}
\beta(k)\, & = \lim_{N\to\infty}\frac{1}{N}\sum_{n=0}^{N-1}
               \frac{\log\bigl(1-\cos(2^{n+1}\pi k)\bigr)}{\log(2)}\\[1mm]
           & =  \int_{0}^{1}\frac{\log\bigl(1-
                  \cos(2 \pi x)\bigr)}{\log(2)} \dd x
           \, = -1 \ts .
\end{aligned}
\]
Such wave numbers $k$ thus do not contribute to the TM measure.

Among the values $k$ for which this is true, we have (possibly up to a
null set) those ones for which the sequence $\bigl( 2^n
k\bigr)_{n\in\NN_{0}}$ is uniformly distributed modulo $1$, none of
which ever visits a dyadic point, and thus never meets the singular
points of $f$.

Note that this argument shows that $\lim_{n\to\infty}f_{n}(k)=0$
pointwise for almost all $k\in\RR$ and thus provides an alternative
proof of the fact that the measure from Eq.~\eqref{eq:TM-Riesz} does
not comprise an absolutely continuous part; compare the Introduction
as well as \cite{Kaku,BG08}.

\emph{Case C.} When $k=\frac{m}{3}$ with $m$ not divisible by $3$, one
finds $f_{n}(k)=(3/2)^{n}$. Since this corresponds to a system (or
sequence) of length $2^{n}$, we have a growth rate of
\[
   \beta(k) \, = \, \frac{\log(3/2)}{\log (2)}\, \approx\, 
   0.584963\, .
\]
The same growth rate applies to all numbers of the form
$k=\frac{m}{2_{\vphantom{I}}^{r} \cdot \, 3}$ with $r \ge 0$ and $m$
not divisible by $3$, because the factor $2^{r}$ in the denominator
has no influence on the asymptotic scaling, due to the structure of
the Riesz product \eqref{eq:TM-Riesz}. Note that the points of this
form are dense in $\RR$, but countable.

Similarly, when $k=\frac{m}{2_{\vphantom{I}}^{r}\cdot\, 5}$ with $r\ge
0$ and $m$ not a multiple of $5$, one finds
\[
     \beta(k)\, = \, \frac{\log(5/4)}{2\ts\log(2)} \, \approx \, 0.160964\, .
\]

\emph{Case D.}  More generally, when ${k=\frac{p}{2_{\vphantom{I}}^{r}q}}$ 
with ${r\ge 0}$, ${q\ge 3}$ odd and $\gcd(p,q)=1$, one can
determine the growth rate explicitly.  Recall that ${U_{\nts q}}
  := {(\ZZ/q\ZZ)^{\times}} = {\{ 1 \le p < q \mid \gcd (p,q)=1 \}}$ is
the unit group of the finite residue class ring $\ZZ/q\ZZ$.  If $S_{q} = {\{
2^n \bmod q \mid n\ge 0 \}}$ is the subgroup of $U_{\nts q}$
generated by the unit $2$, one finds
\begin{equation}\label{eq:betapq}
     \beta (k) \, = \, \frac{1}{ \mathrm{card} (p\ts S_{q})}
     \sum_{n\in pS_{q}} \frac{\log 
     \bigl(1-\cos (2 \pi \frac{n}{q}) \bigr)}{\log (2)}
\end{equation}  
by an elementary calculation. When $q=2m+1$, the integer
$\mathrm{card} (S_{q})$ is the multiplicative order of $2$ mod $q$,
which is sequence \textsf{A\ts 002326} in \cite{Sloane}. 

When $\gcd(p,q)=1$, one has $\mathrm{card}(p\ts S_{q}) =
\mathrm{card}(S_{q})$, even when the set $p\ts S_{q}$ is considered
mod $q$.  Note that formula \eqref{eq:betapq} is written in such a way
that it also holds for all $p$ not divisible by $q$. If $\gcd(p,q)>1$,
the set $p\ts S_{q}$ may be reduced mod $q$, which shows that the
formula consistently gives $\beta(p/q)$ in such cases.

\begin{figure}
\centerline{\includegraphics[width=\columnwidth]{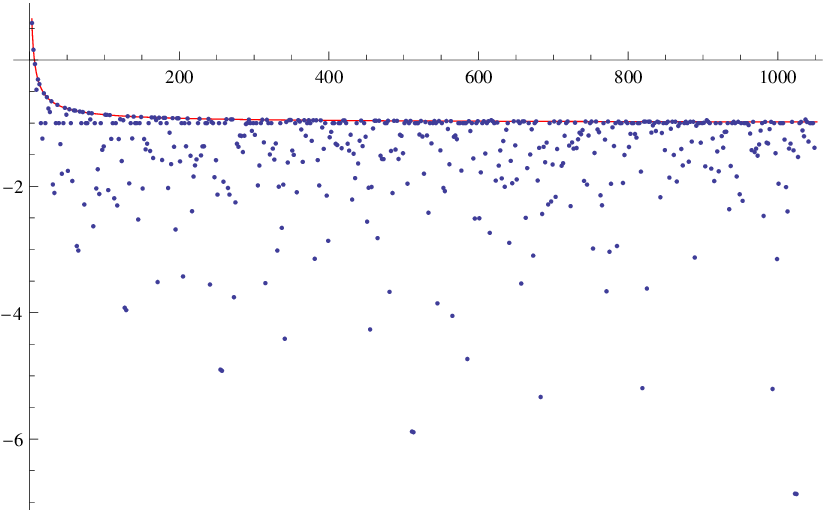}}
\caption{Exponents $\beta(1/q)$ according to Eq.~\eqref{eq:betapq} for
  all odd $3\le q< 1050$. Apart from $q=3$ and $q=5$, the exponents
  are negative. The solid line is the function $g$ from
  Eq.~\eqref{eq:top}. Exponents with large negative values emerge for
  $q=2^{r}\pm 1$.\label{fig:betafig}}
\end{figure}

\emph{Case E.}  
  When $\mathrm{card}(S_{q})=q-1$, Eq.~\eqref{eq:qsum} leads to
  $\beta(1/q)=g(q)$ with
\begin{equation}\label{eq:top}
  g(q) \, = \,  \frac{\log\Bigl(
   \frac{q^{2}}{2_{\vphantom{I}}^{q-1}}\Bigr)}
   {\log\bigl( 2_{\vphantom{I}}^{q-1}\bigr)}  \, = \, 
  \frac{2\ts\log(q)}{(q-1)\log(2)}  -  1\ts .
\end{equation}
For odd $q\ge 3$, the function $g(q)$ is positive precisely for $q=3$
and $q=5$, and negative otherwise; compare Figure~\ref{fig:betafig}.
In fact, also $\beta(1/q)$ seems to be negative for all odd $q\ge 7$,
though this does not hold for general $\beta(p/q)$. Indeed,
$\beta(3/17)>0$, and all positive exponents for odd $7\le q<1000$ are
listed in Table~\ref{qtab}.

More generally, for any odd $q\ge 3$, one obtains (from Case D) the
formula
\begin{equation}
   \frac{1}{q-1} \sum_{1<d|q} \mathrm{card}(S_{d})
   \sum_{p\in U_{\nts d}/S_{d}} \beta\bigl(\tfrac{p}{d}\bigr) 
    \, = \, g(q)\ts .
\end{equation}
Now, M\"{o}bius inversion (with the M\"{o}bius function $\mu$) leads
to
\begin{equation}
  \sum_{p\in U_{\nts q}/S_{q}} \beta\bigl(\tfrac{p}{q}\bigr)
  \, = \, \frac{1}{\mathrm{card}(S_{q})} \sum_{1\ne d|q} 
   \mu\bigl(\tfrac{q}{d}\bigr)\, (d-1)\, g(d)\ts ,
\end{equation}
while it seems difficult to find a simpler formula than
Eq.~\eqref{eq:betapq} for the individual exponents in general.

\emph{Case F.}  As is shown in \cite{CSM} (by way of an explicit
example), there are wave numbers $k$ for which the exponent $\beta(k)$
does not exist. The construction is based on a suitable mixture of
binary expansions for wave numbers with different exponents. Clearly,
there are uncountably many such examples, though they still form a
null set. Here, one can define a `spectrum' of exponents via the
limits of all converging subsequences.

\begin{table}\renewcommand{\arraystretch}{1.3}
  \caption{Wave numbers $k=\frac{p}{q}$ with positive exponents, for all odd
    integers $5<q<1000$. For a given $q$, all $p\in U_{\nts q}/S_{q}$ are
    considered (we choose the smallest element of the set $p\ts S_{q}$ mod $q$
    as representative).\label{qtab}}
\begin{footnotesize}
\begin{tabular}{|@{\,}cl@{\,}|@{\,}cl@{\,}|@{\,}cl@{\,}|@{\,}cl
                @{\,}|@{\,}cl@{\,}|@{\,}cl@{\,}|}\hline
$\frac{p}{q}$ & $\,\beta\bigl(\frac{p}{q}\bigr)$ &
$\frac{p}{q}$ & $\,\beta\bigl(\frac{p}{q}\bigr)$ &
$\frac{p}{q}$ & $\,\beta\bigl(\frac{p}{q}\bigr)$ &
$\frac{p}{q}$ & $\,\beta\bigl(\frac{p}{q}\bigr)$ &
$\frac{p}{q}$ & $\,\beta\bigl(\frac{p}{q}\bigr)$ &
$\frac{p}{q}$ & $\,\beta\bigl(\frac{p}{q}\bigr)$\rule[-6pt]{0pt}{6pt}\\ \hline
 $\frac{3}{17}$    & 0.266 &
 $\frac{25}{117}$  & 0.172 &
 $\frac{37}{255}$  & 0.150 &
 $\frac{65}{381}$  & 0.067 &
 $\frac{47}{565}$  & 0.144 &
 $\frac{65}{771}$  & 0.140 \\
 $\frac{5}{31}$    & 0.272 &
 $\frac{19}{127}$  & 0.108 &
 $\frac{43}{255}$  & 0.318 &
 $\frac{47}{451}$  & 0.127 &
 $\frac{81}{565}$  & 0.113 &
 $\frac{161}{771}$ & 0.140 \\
 $\frac{11}{31}$   & 0.272 &
 $\frac{21}{127}$  & 0.373 &
 $\frac{53}{255}$  & 0.318 &
 $\frac{65}{451}$  & 0.127 &
 $\frac{61}{585}$  & 0.126 &
 $\frac{69}{775}$  & 0.101 \\
 $\frac{5}{33}$    & 0.105 &
 $\frac{27}{127}$  & 0.108 &
 $\frac{91}{255}$  & 0.150 &
 $\frac{67}{455}$  & 0.128 &
 $\frac{97}{585}$  & 0.126 &
 $\frac{83}{775}$  & 0.127 \\
 $\frac{7}{43}$    & 0.267 &
 $\frac{43}{127}$  & 0.373 &
 $\frac{37}{257}$  & 0.049 &
 $\frac{69}{455}$  & 0.128 &
 $\frac{53}{595}$  & 0.031 &
 $\frac{111}{775}$ & 0.127 \\
 $\frac{11}{63}$   & 0.244 &
 $\frac{19}{129}$  & 0.143 &
 $\frac{43}{257}$  & 0.404 &
 $\frac{53}{511}$  & 0.028 &
 $\frac{87}{595}$  & 0.031 &
 $\frac{117}{775}$ & 0.101 \\
 $\frac{13}{63}$   & 0.244 &
 $\frac{11}{151}$  & 0.012 &
 $\frac{45}{257}$  & 0.221 &
 $\frac{75}{511}$  & 0.163 &
 $\frac{51}{601}$  & 0.042 &
 $\frac{57}{785}$  & 0.085 \\
 $\frac{11}{65}$   & 0.350 &
 $\frac{35}{151}$  & 0.012 &
 $\frac{23}{275}$  & 0.117 &
 $\frac{83}{511}$  & 0.239 &
 $\frac{63}{601}$  & 0.042 &
 $\frac{137}{819}$ & 0.359 \\
 $\frac{11}{73}$   & 0.165 &
 $\frac{25}{171}$  & 0.220 &
 $\frac{49}{275}$  & 0.117 &
 $\frac{85}{511}$  & 0.422 &
 $\frac{53}{657}$  & 0.061 &
 $\frac{145}{819}$ & 0.359 \\
 $\frac{13}{73}$   & 0.165 &
 $\frac{19}{185}$  & 0.126 &
 $\frac{25}{331}$  & 0.067 &
 $\frac{87}{511}$  & 0.028 &
 $\frac{101}{657}$ & 0.061 &
 $\frac{67}{825}$  & 0.089 \\
 $\frac{13}{89}$   & 0.229 &
 $\frac{17}{195}$  & 0.001 &
 $\frac{35}{337}$  & 0.149 &
 $\frac{107}{511}$ & 0.239 &
 $\frac{51}{673}$  & 0.038 &
 $\frac{173}{825}$ & 0.089 \\
 $\frac{19}{89}$   & 0.229 &
 $\frac{41}{195}$  & 0.001 &
 $\frac{57}{337}$  & 0.149 &
 $\frac{109}{511}$ & 0.163 &
 $\frac{57}{683}$  & 0.131 &
 $\frac{67}{889}$  & 0.050 \\
 $\frac{9}{91}$    & 0.075 &
 $\frac{17}{205}$  & 0.047 &
 $\frac{49}{341}$  & 0.060 &
 $\frac{171}{511}$ & 0.422 &
 $\frac{71}{683}$  & 0.087 &
 $\frac{95}{889}$  & 0.012 \\
 $\frac{19}{91}$   & 0.075 &
 $\frac{31}{205}$  & 0.183 &
 $\frac{57}{341}$  & 0.369 &
 $\frac{43}{513}$  & 0.033 &
 $\frac{103}{683}$ & 0.179 &
 $\frac{129}{889}$ & 0.012 \\
 $\frac{11}{105}$  & 0.060 &
 $\frac{19}{217}$  & 0.073 &
 $\frac{71}{341}$  & 0.369 &
 $\frac{77}{513}$  & 0.122 &
 $\frac{111}{683}$ & 0.226 &
 $\frac{157}{889}$ & 0.050 \\
 $\frac{17}{105}$  & 0.060 &
 $\frac{37}{217}$  & 0.073 &
 $\frac{73}{341}$  & 0.060 &
 $\frac{83}{513}$  & 0.272 &
 $\frac{113}{683}$ & 0.335 &
 $\frac{83}{993}$  & 0.124 \\
 $\frac{17}{117}$  & 0.172 &
 $\frac{35}{241}$  & 0.194 &
 $\frac{31}{381}$  & 0.067 &
 $\frac{85}{513}$  & 0.343 &
 $\frac{55}{753}$  & 0.054 &
 $\frac{149}{993}$ & 0.172\rule[-5pt]{0pt}{5pt}\\ \hline
\end{tabular}
\end{footnotesize}
\end{table}

\emph{Case G.}  So far, we have identified countably many values of
$k$, for which the scaling exponents can be calculated, while 
(due to Case B) Lebesgue-almost all $k\in\RR$ carry no singular
peak.  The remaining problem is to cope with the uncountably many wave
numbers (of zero Lebesgue measure) that belong to the supporting set
of the TM measure and may possess well-defined exponents.

The existence of such numbers can be understood via Diophantine
approximation. Again, it is useful to start with the binary expansion
of a wave number $k$, and then modify it in a suitable way. Consider
first the example
\[
    k \, = \, \tfrac{1}{3} \, = \, 0.0101010101\ldots
\]
If we now switch the binary digits at positions $2^r$, with $r\in\NN$,
we obtain a different wave number $k'$ that is irrational but
nevertheless still has the same scaling exponent $\beta$ as
$k=\frac{1}{3}$, as longer and longer stretches of the binary
expansion of $k'$ agree with that of $k$. Clearly, via similar
modifications, we can obtain uncountably many distinct irrational
numbers with $\beta = \beta(1/3)$.

The same strategy works for all other rational wave numbers $k$, and
underlies the nature of the TM measure. In particular, this explains
the existence of uncountably many `singular peaks', which together (in
view of Case~B) still form a Lebesgue null set. These scaling
exponents are accessible via our above arguments. It would be interesting
to see whether one can go any further with the pointwise analysis.

To continue, it would be natural to augment the above results by a
proper analysis of the full TM measure, along the lines of
\cite{GL,Wolny,Zaks,ZPK1,ZPK2} and related works. Also, a better
understanding of the set of wave numbers for which the exponent $\beta
(k)$ is well-defined would be welcome.

\section{Concluding remarks}

An analogous approach works for all measures of the form of a classic
Riesz product. In particular, the generalised Thue--Morse sequences
from \cite{BGG} can be analysed along these lines; compare also
\cite{Kea}. Likewise, the choice of different interval lengths is
possible, though technically more complicated; compare \cite{Wolny}
for some examples.

Higher-dimensional examples with purely singular continuous spectrum,
such as the squiral tiling \cite{squiral} or similar bijective block
substitutions \cite{Nat}, may still lead to classic Riesz products,
though they are now in more than one variable, and the analysis is
hence more involved.  Nevertheless, the scaling analysis will still
lead to a better understanding of such measures.

\section*{Acknowledgements}

We thank Gerhard Keller, Marc Ke{\ss}eb\"{o}hmer and Tanja Schindler
for interesting discussions and comments.  This work was supported by
the German Research Foundation (DFG) within the CRC~701.


\begin{thebibliography}{99}

\bibitem{AS}
J.-P.~Allouche and J.~Shallit,
\textit{Automatic Sequences:\ Theory,\ Applications,\ Generalizations},
Cambridge University Press, Cambridge (2003).

\bibitem{BG08}
M.~Baake and U.~Grimm,
The singular continuous diffraction measure of the Thue--Morse chain,
\textit{J.\ Phys.\ A:\ Math.\ Theor.} \textbf{41}, 422001 (2008); 
\texttt{arXiv:0809.0580}.

\bibitem{squiral}
M.~Baake and U.~Grimm,
Squirals and beyond:\ Substitution tilings with singular
continuous spectrum, 
\textit{Erg.\ Th.\ \& Dynam.\ Syst.}, in press;
\texttt{arXiv:1205.1384}.

\bibitem{tao}
M.~Baake and U.~Grimm,
\textit{Aperiodic Order. Vol.~$1$: A Mathematical Invitation},
Cambridge University Press, Cambridge (2013).

\bibitem{BGG}
M.~Baake, F.~G\"{a}hler and U.~Grimm,
Spectral and topological properties of a family of 
generalised Thue--Morse sequences,
\textit{J.\ Math.\ Phys.} \textbf{53}, 032701 (2012);
\texttt{arXiv:1201.1423}.

\bibitem{CSM}
Z.~Cheng, R.~Savit and R.~Merlin,
Structure and electronic properties of Thue--Morse lattices,
\textit{Phys.\ Rev.\ B} \textbf{37}, 4375--4382 (1988).

\bibitem{Nat}
N.P.~Frank,
Multi-dimensional constant-length substitution sequences,
\textit{Topol.\ Appl.} \textbf{152}, 44--69 (2005).

\bibitem{GL} 
C.~Godr\'{e}che and J.M.~Luck, 
Multifractal analysis in reciprocal space and the nature of the
Fourier transform of self-similar structures,
\textit{J.\ Phys.\ A:\ Math.\ Gen.} \textbf{23}, 3769--3797 (1990).

\bibitem{HL}
G.H.~Hardy and J.E.~Littlewood,
Some problems of Diophantine approximation,
\textit{Acta Math.} \textbf{37}, 155--191 (1914).

\bibitem{Kac}
M.~Kac,
On the distribution of values of sums of the type
$\Sigma f(2^k t)$,
\textit{Ann.\ Math.} \textbf{47}, 33--49 (1946).

\bibitem{Kaku} 
S.~Kakutani,
Strictly ergodic symbolic dynamical systems, in:
\textit{Proc.\ 6th Berkeley Symposium on Math.\ Statistics
and Probability} eds L.M.~LeCam, J.~Neyman and 
E.L.~Scott, Univ.\ of California Press, Berkeley (1972),
pp.\ 319--326.

\bibitem{Kea}
M.~Keane,
Generalized Morse sequences,
\textit{Z.\ Wahrscheinlichkeitsth.\ verw.\ Geb.}
\textbf{10}, 335--353 (1968).

\bibitem{KN}
L.~Kuipers and H.~Niederreiter,
\textit{Uniform Distribution of Sequences},
Wiley, New York (1974); reprint Dover, New York (2006).

\bibitem{Mah} 
K.~Mahler,
The spectrum of an array and its application to the study of the
translation properties of a simple class of arithmetical functions.
Part II:\ On the translation properties of a simple class of arithmetical
functions,
\textit{J.\ Math.\ Massachusetts} \textbf{6}, 158--163 (1927). 

\bibitem{Q}
M.~Queff\'{e}lec,
\textit{Substitution Dynamical Systems -- Spectral Analysis},
LNM 1294, 2nd ed., Springer, Berlin (2010).

\bibitem{Sloane}
N.J.A.S.~Sloane,
\textit{The On-Line Encyclopedia of Integer Sequences},
available at \texttt{http://oeis.org/}.

\bibitem{ME}
A.C.D.~van Enter and J.~Mi\c{e}kisz,
How should one define a weak crystal?
\textit{J.\ Stat.\ Phys.} \textbf{66}, 1147--1153 (1992).

\bibitem{Wie} 
N.~Wiener,
The spectrum of an array and its application to the study of the
translation properties of a simple class of arithmetical functions.
Part I:\ The spectrum of an array,
\textit{J.\ Math.\ Massachusetts} \textbf{6}, 145--157 (1927).

\bibitem{Withers}
R.L.~Withers,
Disorder, structured diffuse scattering and the
transmission electron microscope,
\textit{Z.\ Krist.} \textbf{220}, 1027--1034 (2005).
 
\bibitem{Wolny} 
J.~Wolny, A.~Wn\c{e}k and J.-L.\ Verger-Gaugry,
Fractal behaviour of diffraction pattern of Thue--Morse sequence
\textit{J.\ Comput.\ Phys.} \textbf{163}, 313--327 (2000).

\bibitem{Zaks}
M.A.~Zaks,
On the dimensions of the spectral measure of
symmetric binary substitutions,
\textit{J.\ Phys.\ A:\ Math.\ Gen.} \textbf{35}, 5833--5841 (2002).

\bibitem{ZPK1}
M.A.~Zaks, A.S.~Pikovsky and J.~Kurths,
On the correlation dimension of the spectral measure for the Thue--Morse 
sequence,
\textit{J.\ Stat.\ Phys.} \textbf{88}, 1387--1392 (1997).

\bibitem{ZPK2}
M.A.~Zaks, A.S.~Pikovsky and J.~Kurths,
On the generalized dimensions for the Fourier spectrum of the Thue--Morse 
sequence,
\textit{J.\ Phys.\ A:\ Math.\ Gen.} \textbf{32},  1523--1530 (1999).

\bibitem{Z}
A.~Zygmund,
\textit{Trigonometric Series},
3rd ed., Cambridge University Press, Cambridge (2002).

\end{thebibliography}
\end{document}